\documentclass[conference]{IEEEtran}
\usepackage{amsmath}
\usepackage{amssymb}
\usepackage[dvips]{graphicx}
\usepackage{epsfig}

\newcommand{\X}{\mathbf{X}}

\newcommand{\z}{\mathbf{z}}
\newcommand{\h}{\mathbf{h}}

\newcommand{\hh}{\mathbf{H}}
\newcommand{\w}{\mathbf{w}}

\newcommand{\I}{\mathbf{I}}

\newcommand{\bv}{\mathbf{v}}

\newcommand{\bu}{\mathbf{u}}

\newtheorem{lemma:bitenergylow}{Lemma}
\newtheorem{lemma:bitenergyhigh}[lemma:bitenergylow]{Lemma}

\newtheorem{prop:asympcap}{Theorem}
\newtheorem{prop:flashminbitenergy}[prop:asympcap]{Theorem}
\newtheorem{prop:flashbitenergy}[prop:asympcap]{Theorem}
\newtheorem{prop:pasympcap}[prop:asympcap]{Theorem}

\begin{document}


\title{Collaborative Relay Beamforming for Secrecy}



%
\author{\authorblockN{Junwei Zhang and Mustafa Cenk Gursoy}
\authorblockA{Department of Electrical Engineering\\
University of Nebraska-Lincoln, Lincoln, NE 68588\\ Email:
junwei.zhang@huskers.unl.edu, gursoy@engr.unl.edu}}

\maketitle
\begin{abstract}\footnote{This work was supported by the National Science Foundation under Grant CCF -- 0546384 (CAREER).}
In this paper, collaborative use of relays to form a
beamforming system with the aid of perfect channel state
information (CSI) and to provide physical-layer security is investigated. In
particular, a decode-and-forward-based relay beamforming
design subject to total and
individual relay power constraints is  studied with the goal of maximizing the secrecy rate. It is shown that the total power constraint leads to a closed-form solution. The design under individual relay power
constraints is formulated as an optimization problem which is shown to be easily solved using two different approaches, namely semidefinite programming and second-order cone programming. Furthermore, a simplified and
suboptimal technique which reduces the computation complexity under
individual power constraints is presented. 
\end{abstract}

\section{introduction}
The open nature of wireless communications allows for the signals to
be received by all users within the communication range. Thus,
wireless communication is vulnerable to eavesdropping. This problem
was first studied from an information-theoretic perspective in \cite{wyner} where Wyner proposed a  wiretap
channel model. Wyner showed that secure communication is possible
without sharing a secret key if the eavesdropper's channel is a
degraded version of the main chain.  Later, Wyner's result was
extended to the Gaussian channel in \cite{cheong} and recently to
fading channels in \cite{Gopala}. In addition to the single antenna
case, secrecy in multi-antenna models is addressed in \cite{shafiee}
 and \cite{khisti}.  Liu  \emph{et al.} \cite{Liu} presented inner and
outer bounds on secrecy capacity regions for broadcast and
interference channels. The secrecy capacity of  multi-antenna
broadcasting channel is obtained in \cite{Liu1}. Additionally,
it's well known that that users can cooperate to form a distributed
multi-antenna system by relaying. Cooperative relaying under secrecy
constraints was recently studied in \cite{dong}--\cite{aggarwal} .

In \cite{dong}, a decode-and-forward (DF) based cooperative protocol
is considered, and a beamforming system is designed
for secrecy capacity maximization or transmit power minimization.
However, in this work, only total relay power constraint is imposed.  In
this paper, we extend the analysis to both total and individual
constraints. Under individual power constraints, although analytical solutions are not available, we provide an
optimization framework. We use the semidefinite relaxation (SDR)
approach to approximate the problem as a convex semidefinite
programming (SDP) problem which can be solved efficiently. We also
provide an alternative method by formatting the original
optimization problem as a convex second-order cone programming (SOCP)
problem that can be efficiently solved by interior point methods.
Finally, we describe a simplified suboptimal beamformer design under
individual power constraints. Simulation results are given to compare
the performances of different beamforming methods.

\section{Channel Model}

\begin{figure}
\begin{center}
\includegraphics[width = 0.5\textwidth]{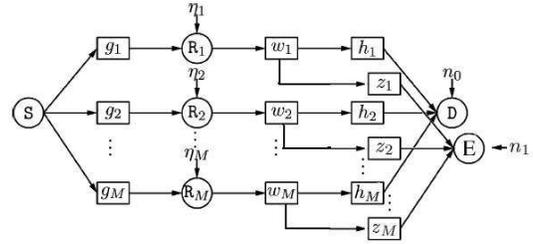}
\caption{Channel Model} \label{fig:channel}
\end{center}
\end{figure}

We consider a communication channel with a source $S$, a destination
$D$, an eavesdropper $E$, and $M$ relays $\{R_m\}_{m=1}^M$ as depicted
in Figure.\ref{fig:channel}. We assume that there is no direct link
between $S$ and $D$. We also assume that relays work synchronously
by multiplying the signals to be transmitted with complex weights $\{w_m\}$ to produce a virtual beam
point to destination and eavesdropper. We denote the channel fading coefficient
between $S$ and $R_m$ as $g_m\in \mathbb{C}$, the fading coefficient between
$R_m$ and $D$ as $h_m\in \mathbb{C}$, and the fading coefficient between $R_m$
and $E$ as $z_m\in \mathbb{C}$. In this model, the source $S$ tries to transmit
confidential messages to $D$ with the help of the relays  while keeping the
eavesdropper $E$ ignorant of the information.

It's obvious that our channel is a two-hop relay network. In the first
hop, the source $S$ transmits $x_s$ to relays with power
$E[|x_s|^2]=P_s$. The received signal at $R_m$ is given by
\begin{align}
y_{r,m}=g_m x_s+\eta_m
\end{align}
where $\eta_m$ is the background noise that has a Gaussian distribution with zero mean and variance of $N_m$.

In the second hop, we employ decode and forward transmission scheme.
In this scheme, each relay $R_m$ first decodes the message $x_s$ and
normalizes it as $ x_s'=x_s/P_s$. Subsequently, the normalized message is
multiplied by the weight factor $w_m$ to generate the transmitted signal $x_r=w_m x_s'$. The output power of
each relay $R_m$ is given by
\begin{align}
E[|x_r|^2]=E[|w_m  x_s'|^2]=|w_m|^2
\end{align}
The received signals at the destination $D$ and eavesdropper $E$ are
the superpositions of the signals transmitted from the relays. These signals can be expressed, respectively, as
\begin{align}
y_d&=\sum_{m=1}^M h_m w_m  x_s' +n_0=\h^\dagger \w x_s' +n_0, \quad \text{and} \\
y_e&=\sum_{m=1}^M z_m w_m  x_s' +n_1 =\mathbf{z}^\dagger \w x_s' +n_1
\end{align}
where $n_0$ and $n_1$ are the Gaussian background noise components at $D$ and $E$, respectively, with zero mean
and variance $N_0$.  Additionally, we have defined $\mathbf{h}=[h_1^*,....h_M^*]^T,
\mathbf{z}=[z_1^*,....z_M^*]^T$, and $\w=[w_1,...w_M]^T$ where superscript $*$
denotes conjugate operation, and $(\cdot)^T$ and $(\cdot)^\dagger $ denote the
transpose and conjugate transpose, respectively, of a matrix or vector. The metrics of interest
are the received SNR levels at $D$ and $E$, which are given by
\begin{align}
\Gamma_d&=\frac{|\sum_{m=1}^M h_m w_m|^2}{N_0} \quad \text{and}\\
\Gamma_e&=\frac{|\sum_{m=1}^M z_m w_m|^2}{N_0}.
\end{align}
It has been proved that the secrecy rate $R_s$ over the channel between the relays and destination is
\begin{align}
R_s&=I(x_s;y_d)-I(x_s;y_e)\\
&=\log(1+\Gamma_d)-\log(1+\Gamma_e)\\
&=\log\left(\frac{N_0+|\sum_{m=1}^M h_m w_m|^2}{N_0+|\sum_{m=1}^M z_m
w_m|^2}\right) \label{eq:secrecyrate}
\end{align}
where $I(\cdot;\cdot)$ denotes the mutual information. 
In this paper, we address the joint optimization of $\{w_m\}$ with the aid perfect CSI and hence identify the optimum
collaborative relay beamforming (CRB) direction that maximizes the secrecy rate given in (\ref{eq:secrecyrate}).

\section{Optimal Beamforming 
under total power constraints}

In this section, we consider a total relay
power constraint in the following form: $||\w||^2=\w^\dagger \w\leq P_T$. The optimization problem can now be formulated as follows:
\begin{align}\label{dftotal}
C_s(\h,\z,P_T)&=\max_{\w^\dagger \w\leq P_T}\log\left(\frac{N_0+|\sum_{m=1}^M h_m
w_m|^2}{N_0+|\sum_{m=1}^M z_m
w_m|^2}\right) \nonumber\\
&=\log  \max_{\w^\dagger \w \leq P_T}\frac{N_0+|\sum_{m=1}^M h_m
w_m|^2}{N_0+|\sum_{m=1}^M z_m w_m|^2}\\
&=\log  \max_{\w^\dagger \w\leq P_T} \frac{\w^\dagger(\frac{N_0}{P_T}\I+\h
\h^\dagger)\w}{\w^\dagger(\frac{N_0}{P_T}\I+\z \z^\dagger)\w}\\
&=\log  \max_{\w^\dagger \w\leq P_T} \frac{\w^\dagger(N_0\I+P_T\h
\h^\dagger)\w}{\w^\dagger(N_0 \I+ P_T\z \z^\dagger)\w}\\
&=\log \lambda_{\max}(N_0\I+P_T\h \h^\dagger,N_0\I+P_T\z\z^\dagger) \label{eq:maxsecrecyrate}
\end{align}
where $\lambda_{\max}(\mathbf{A},\mathbf{B})$ is the largest
generalized eigenvalue of the matrix pair $(\mathbf{A},\mathbf{B})$. Hence, the maximum secrecy rate in (\ref{eq:maxsecrecyrate}) is achieved by the optimum beamforming weights given by
\begin{align}\label{woptdfto}
\w_{opt}=\varsigma \bu
\end{align}
where $\bu$ is the eigenvector that corresponds to
$\lambda_{\max}(N_0\I+P_T\h \h^\dagger,N_0\I+P_T\z\z^\dagger)$  and $\varsigma$ is chosen to ensure $\w_{opt}^\dagger \w_{opt}
=P_T$.

Note that in the first-hop of the channel model, the maximum rate we can achieve is
\begin{align}
C_1=\min_{m=1,\ldots,M} \log\left(1+\frac{|g_m|^2P_s}{N_m}\right)
\end{align}
Thus, the overall secrecy rate is
\begin{align}
C_{dof,s}=\min(C_1, C_s)
\end{align}
Next, we provide some remarks on the performance of
collaborative relay beamforming in the high and low SNR regimes. For
simplicity, we assume in the following that the noise variances at the destination and
eavesdropper are $N_0=1$.

\subsection{High SNR Regime} 

In the high SNR scenario, where
both $P_s, P_T \to \infty  $ and $ \lim \frac{P_s}{P_T}=1$, we can
easily see that
\begin{align}
\lim_{P_s \to \infty} C_1=\log P_s \min_{m=1}^M \log (|g_m|^2/N_m)
\end{align}
>From the result in \cite{khisti}, we can see that
\begin{align}
\lim_{P_T \to \infty} C_s&=\log
\lambda_{\max}(\h\h^\dagger,\z\z^\dagger)\\
&=\log P_T+\log(\max_{\tilde{\psi}} |\h^\dagger
\tilde{\psi}|^2)\label{h1}
\end{align}
$\tilde{\psi}$ unit vector on the null space of $\z^\dagger$.

Based on these observation, at high SNR, we can choose the
beamforming vectors to lie in the null spaces of the eavesdropper's
channel vector. That is $|\sum_{m=1}^M z_m w_m|^2=\z^\dagger \w=0$.
In this case, the eavesdropper can not receive any data transmission
from relays, the secrecy constraint is automatically guarantied. No
secrecy coding are needed at relays. Now as $P_T \to\infty  $ the
weights optimization problem becomes
\begin{align}
\max_{\w^\dagger \w\leq  P_T}&\log \left(\left|\sum_{m=1}^M
h_m w_m\right|^2\right) ~~~~~s.t ~~\z^\dagger \w=0 \label{dbrd}\\
&= \log( P_T)+\log(\max_{\tilde{\psi}} |\h^\dagger
\tilde{\psi}|^2)\label{h2}
\end{align}
since (\ref{h2}) is identical as (\ref{h1}), the null space
beamforming is optimal at high SNR. Furthermore, the optimal null
space beamforming vector can be obtained explicitly.

Due to the null space constraint, we can write $\w=\hh_{z}^\bot
\bv$, where $\hh_{z}^\bot$ denote the projection matrix onto the
null space $\z^\dagger$. Specifically, the column of $\hh_{z}^\bot$
are orthonormal vector which composed the basis of null spaces of
$\z^\dagger$. In our case,  $\hh_{z}^\bot$ is a $M*(M-1)$ matrix.
The power constraint $\w^\dagger \w= \bv^\dagger
{\hh_{z}^\bot}^\dagger \hh_{z}^\bot \bv=\bv^\dagger \bv\leq P_T$.
Then the weights optimization problem  can be written as,

\begin{align}\label{dftotal}
\max_{\w^\dagger \w\leq  P_T}&\log
\left(1+|\sum_{m=1}^M h_m w_m|^2\right)\\
&=\log \left(1+\max_{\w^\dagger \w\leq  P_T} (\w^\dagger \h \h^\dagger \w)\right)\\
&=\log \left(1+\max_{\bv^\dagger \bv\leq  P_T}
(\bv^\dagger{\hh_{z}^\bot}^\dagger \h \h^\dagger
{\hh_{z}^\bot}\bv)\right)\\
&=\log \left(1+P_T\lambda_{\max}
({\hh_{z}^\bot}^\dagger \h \h^\dagger {\hh_{z}^\bot})\right)\label{a}\\
&=\log \left(1+P_T \h^\dagger {\hh_{z}^\bot}{\hh_{z}^\bot}^\dagger
\h\right)\label{b}
\end{align}
the optimum null space beamforming weights $\w$ is
\begin{align}\label{woptdfto}
\w_{opt,n}=\hh_{z}^\bot \bv=\varsigma_1 \hh_{z}^\bot
{\hh_{z}^\bot}^\dagger \h
\end{align}
$\varsigma_1$ is constant satisfied power constraint.

\subsection{Low SNR Regime}

In the low SNR scenario ,where  both $P_s, P_T \to 0  $ and $\lim
\frac{P_s}{P_T}=1$.We can  see that
\begin{align}
\lim_{P_s \to 0} C_1=P_s  \min_{m=1}^M \frac{|g_m|^2}{N_m}
\end{align}

\begin{align}
\lim_{P_s \to 0} C_s=P_T \lambda_{\max}(\h \h^ \dagger-\z \z^\dagger)
\end{align}

Thus, in the low SNR regime, the direction of optimal CRB vector
approaches the regular eigenvector to the largest regular eigenvalue
of $\h \h^ \dagger-\z \z^\dagger$.

%

\section{Optimal Beamforming under Individual power constraints}

In a multiuser network such as the relay system we study in this
paper, it becomes necessary to consider individual power constraint
at the relays, which can be reflected by having $|w_m|^2 \leq p_m
\forall m $ or $|\w| \leq \mathbf{p}$ where we have $|\cdot|^2$ to
denote element-wise norm-square operation and $\mathbf{p}$ is a
column vector containing the element $\{p_m\}$. In what follows, the
problem of interest will be maximization the term inside $\log $
function of $R_s$
\begin{align}
&\max_{|\w|^2 \leq \mathbf{p}} \frac{N_0+|\sum_{m=1}^M h_m
w_m|^2}{N_0+|\sum_{m=1}^M z_m w_m|^2} \label{optind1}\\
&=\max_{|\w|^2 \leq \mathbf{p}} \frac{N_0+\w^\dagger \h \h^\dagger
\w}{N_0+\w^\dagger \z \z^\dagger \w} \label{optind}
\end{align}
\subsection{Semidefinite Relaxation (SDR) Approach}
We first consider a semidefinite program method similar to that in
\cite{luo}. Using the definition $\X\ \triangleq \w \w ^\dagger$,
the optimization problem (\ref{optind}) can be write as
\begin{align}
\max_{\X}~~ &\frac{N_0+Tr(\h \h^\dagger \X)}{N_0+Tr(\z \z^\dagger
\X)}
\label{SDR1} \\
s.t ~~& ~~diag(\X)\leq \mathbf{p} \nonumber \\
&~~ rank ~~\X=1,~~~ and ~~~\X\succeq 0 \nonumber
\end{align}
or equivalently as
\begin{align}
\max_{\X, t} &~~~t \label{SDR} \\
s.t &~~ tr(\X(\h\h^\dagger- t\z\z^\dagger))\geq N_0(t-1),\nonumber\\
~~& ~~diag(\X)\leq \mathbf{p}, \nonumber \\
&~~ rank ~~\X=1,~~~ and ~~~\X\succeq 0 \nonumber
\end{align}
 Where $tr(\cdot)$ represent the trace of a matrix
and $\X\succeq 0$ means that $\X$  is a symmetric positive
semi-definite matrix. The optimization problem in (\ref{SDR}) is not
convex and may not be easily solved. Let us then ignore the rank
constraint in (\ref{SDR}). That is, using a semi-definite relaxation
(SDR), we aim to solve the following optimization problem:
\begin{align}
&\max_{\X, t} ~~~t \label{SDR2} \\
&s.t ~~ tr(\X(\h\h^\dagger- t\z\z^\dagger))\geq N_0(t-1),\nonumber\\
&and ~~diag(\X)\leq \mathbf{p}, ~~~ and ~~~\X \succeq 0 \nonumber
\end{align}
 If the matrix $\X_{opt}$ obained by solving the optimization problem in
(\ref{SDR2}) happens to be rank one, the its principal component
will be the optimal solution to the original problem. Note that the
optimization problem in (\ref{SDR2}) is quasiconvex. In fact, for
any value of t, the feasible set in (\ref{SDR2}) is convex.  Let
$t_{\max}$ be the maximum value of t obtained by solving the
optimization problem  (\ref{SDR2}). If, for any given $t$, the
convex feasibility problem
\begin{align}
&find~~~~\X \label{SDR3}\\
&such~~that ~ tr(\X(\h\h^\dagger- t\z\z^\dagger))\geq N_0(t-1),\nonumber\\
&and ~~diag(\X)\leq \mathbf{p}, ~~~ and ~~~\X \succeq 0 \nonumber
\end{align}

is feasible, then we have $t_{\max}\geq t$. Conversely, if the convex
feasibility optimization problem (\ref{SDR3}) is not feasible, then
we conclude $t_{\max} <t$. Therefore, we can check whether the
optimal value $t_{\max}$ of the quasiconvex optimization problem in
(\ref{SDR2}) is smaller than or greater than a given value t by
solving the convex feasibility problem (\ref{SDR3}). If the convex
feasibility problem (\ref{SDR3}) is feasible then we know
$t_{\max}\geq t$. If the convex feasibility problem (\ref{SDR3}) is
infeasible, then we know that  $t_{\max}< t$. Based on this
observation, we can use a simple  bisection algorithm to solve the
quasiconvex optimization problem (\ref{SDR2}) by solving a convex
feasibility problem (\ref{SDR3})at each step. We assume that the
problem is feasible, and start with an interval $[l ,u]$ known to
contain the optimal value $t_{\max}$. We then solve the convex
feasibility problem at its midpoint $t = (l + u)/2$, to determine
whether the optimal value is larger or smaller than $t$. We update
the interval accordingly to obtain a new interval. That is, if t is
feasible, then we set $l = t$, otherwise, we choose $u = t$ and
solve the convex feasibility problem  again. This procedure is
repeated until the width of the interval is smaller than the given
threshold. Once the maximum feasible value for $ t_{\max}$ is
obtained, one can solve
\begin{align}
&\min{\X} ~~~tr(\X) \label{SDR4} \\
&s.t ~~ tr(\X(\h\h^\dagger- t_{\max}\z\z^\dagger))\geq N_0(t_{\max}-1),\nonumber\\
&and ~~diag(\X)\leq \mathbf{p}, ~~~ and ~~~\X \succeq 0 \nonumber
\end{align}
to get the solution $\X_{opt}$.(\ref{SDR4}) is a convex problem
which can be solved efficiently using interior-point based method.

To solve the convex feasibility problem, one can use the
well-studied interior point based methods. We use the  well
developed interior point method based package SeDuMi
\cite{sedumi}that produces a feasibility certificate if the problem
is feasible and its popular interface  Yalmip \cite{yalmip}. In
semi-definite relaxation, the solution may not be rank one in
general. Interestingly, in our extensive simulation results, we have
never encountered a case where the solution $\X_{opt}$ to the SDP
problem has a rank higher than one. In fact, there is  always  a
rank one optimal solution exists for our problem which we will
explain later. Then we can obtain our optimal beamforming vector
from principal component of the optimal solution $\X_{opt}$.

\subsection{Second-order Cone Program (SOCP) Approach} The reason
that this SDR method is optimal for above problem is that we can
reformulate the problem as the second order cone problem. Thus, we
have another way of solving it. The optimization
problem(\ref{optind1}) is equivalent  to
\begin{align}
&\max_{\w, t} ~~~~ t \label{socp}\\
&s.t\ \frac{N_0+|\h^\dagger \w|^2}{N_0+|\z^\dagger \w|^2}\geq t \label{socpt}  \\
&and~~~~|\w|^2\leq \mathbf{p}\nonumber
\end{align}
Observe that an arbitrary phase rotation can be added to the
beamforming vector without affecting the constraint (\ref{socpt}).
Thus, $\h^\dagger \w$ can be chosen to be real without loss of
generality. Note (\ref{socpt}) can be write as
\begin{align}
\frac{1}{t}|\h^\dagger \w|^2\geq \left|\left| \begin{array}{ccc}
\z^\dagger \w\\
 \sqrt{1-\frac{1}{t} N_0} \\
\end{array}\right |\right|^2
\end{align}
because $\h^\dagger \w$ can be assume to be real, we may take the
square root of the above equation. The constraint becomes a
second-order cone constraint, which is convex. The optimization
problem now becomes
\begin{align}
&\max_{\w, t} ~~~~ t \label{socp1}\\
&s.t ~~\sqrt{\frac{1}{t}}\h^\dagger \w \geq \left|\left|
\begin{array}{ccc}
\z^\dagger \w\\
 \sqrt{1-\frac{1}{t} N_0} \\
\end{array}\right |\right|
&and~~~~|\w|^2\leq \mathbf{p}\nonumber
\end{align}

As described in  the SDR approach, the optimal solution of
(\ref{socp1}) can be obtained by repeatedly check feasibility  using
a bisection search over $t$ with the aid of well studied interior
point method for second order cone program. Again, we use SeduMi
together with Yalmip in our simulation. Once, the maximum feasible
value  $t_{\max}$ is obtained, we can then solve following second
order cone problem (SOCP) to obtain optimum beamforming vector.

\begin{align}
&\min_{\w} ~~~~ ||\w||^2 \label{socp1}\\
&s.t ~~\sqrt{\frac{1}{t_{\max}}}\h^\dagger \w \geq \left|\left|
\begin{array}{ccc}
\z^\dagger \w \\
 \sqrt{1-\frac{1}{t_{\max}} N_0} \\
\end{array}\right |\right|
&and~~~~|\w|^2\leq \mathbf{p}\nonumber
\end{align}

Thus, we can get the secrecy capacity $C_{s,ind}$ for the second hop
relay beamforming system under individual power constraint by the
above two numerical optimization method. Then combined with first
hop source relay link secrecy capacity $C_1$. Secrecy capacity of
decode and forward CRB system is $ C_{dof,ind}=\min(C_1, C_{s,ind})$

\subsection{Simplified Suboptimal Design} 

As was shown above, design
the beamformer with individual relay power constraints requires an
iterative procedure where, at each step, a convex feasibility
problem is solved. We now try to get a suboptimal beamforming vector
without significant computation complexity.
%

we choose simplified  beamformer as $\mathbf{v}=\theta \w_{opt}$
where $w_{opt}$ is given by (\ref{woptdfto}) with
$||w_{opt}||^2=P_T=\sum p_i$, and
 we choose
\begin{align}
\theta=\frac{1}{|w_{opt,k}|/\sqrt{p_k}}
\end{align}
where $w_{opt,k}, p_k$ is the $k$th entry of $w_{opt}, \mathbf{p}$
respectively and
\begin{align}
k=\arg \max_{1\leq i\leq M} \frac{|w_{opt,i}|^2}{p_i}
\end{align}
Then,substitute this beamformer  $\mathbf{v}$ into (\ref{optind}) to
get the achievable suboptimal rate under individual power
constraint.

\section{Simulation Results}

In our simulations we focus on the performance of $2nd$ hop secrecy
capacity since the main concern of this paper is the design  of the
CRB system in second hop.  We assume  rayleigh distributed
$\{g_m\}$, $\{h_m\}, \{z_m\}$ is complex circularly symmetric
gaussian with zero mean and variance $\sigma_g^2$, $\sigma_h^2$,and
$\sigma_z^2$ respectively.

\begin{figure}
\begin{center}
\includegraphics[width = 0.45\textwidth]{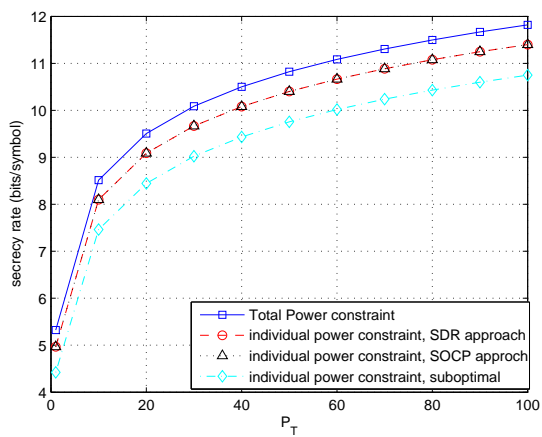}
\caption{2nd hop secrecy rate vs. the total relay transmit power
$P_T$ for different cases, eavesdropper has a weaker channel }
\label{fig:indviDF}
\end{center}
\end{figure}

\begin{figure}
\begin{center}
\includegraphics[width = 0.45\textwidth]{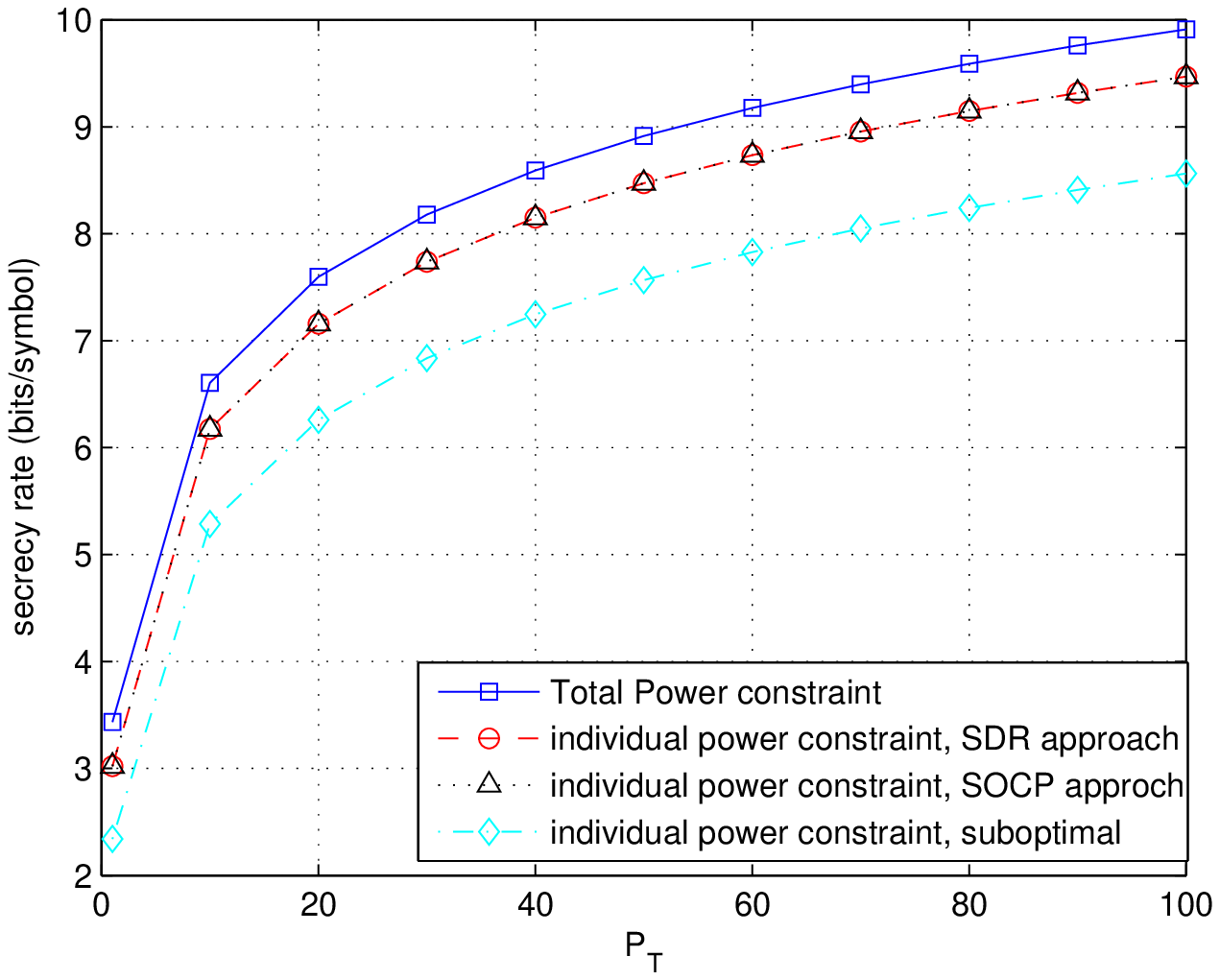}
\caption{2nd hop secrecy rate vs. the total relay  transmit power
$P_T$ for different cases, eavesdropper has a stronger channel }
\label{fig:indviDF1}
\end{center}
\end{figure}

In Fig.\ref{fig:indviDF}  Fig.\ref{fig:indviDF1}, we plotted the
second hop secrecy rate, which is the maximum secrecy rate our
collaborative relay beamforming system can supported for both total
and individual relay power constraint. For individual relay power
constraint, we consider the relays to have the same power budgets:
$p_i=\frac{P_T}{M}$. Specifically, in  Fig.\ref{fig:indviDF}, we
have $\sigma_h=3$, $\sigma_z=1$, $N_0=1$ and $M=5$. In this case,
the legitimate user has a stronger channel. In
Fig.\ref{fig:indviDF1}, we only change $\sigma_h=1$, $\sigma_z=2$
that is the eavesdropper has a stronger channel. Our CRB system can
achieve secrecy transmission even the eavesdropper's has a favorable
channel condition.  As can be seen from
 figures, the secrecy rate under total transmit power and that
under individual relay power are very close to each other. And our
two different optimization approach gives nearly the same result. It
also can be seen that under individual power constraint, the simple
suboptimal method suffers a constant loss as compared to SDR or SOCP
based optimal value.

\begin{figure}
\begin{center}
\includegraphics[width = 0.45\textwidth]{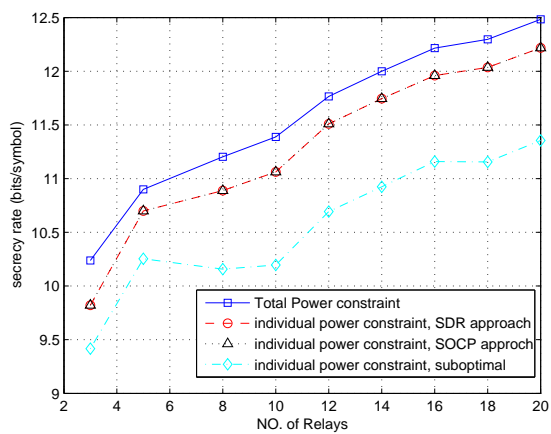}
\caption{2nd hop secrecy rate vs. Number of relays for different
cases} \label{fig:indviDFm}
\end{center}
\end{figure}

In Fig.\ref{fig:indviDFm}, we fix the relay total transmitting power
as $P_T=10$, and vary the number of collaborative relays. Other
parameter is the same as Fig.\ref{fig:indviDF1}. We can see
increasing the $M$ will increasing the secrecy capacity for both
total and individual power constraint. We also see that in some
situation, increasing $M$ will degrade the performance for our
simplified suboptimal beamformer.

\end{document}